\documentclass[onecolumn,draftcls,romanappendices]{IEEEtran}
\usepackage{mathpple}
\usepackage{times}
\usepackage{amsthm}
\usepackage{amsmath}
\usepackage{amssymb}  
\usepackage[mathscr]{eucal}
\usepackage{cite}     
\usepackage{comment}  
\usepackage{upref}
\usepackage{amsfonts}
\usepackage{verbatim}
\usepackage[dvipsnames,usenames]{color}
\usepackage{enumerate}
\usepackage{graphicx}
\usepackage{latexsym}
\usepackage{bm}
\usepackage{multirow}
\usepackage{dsfont}
\usepackage{amsthm,xpatch}
\usepackage{mathtools}
\usepackage{bbm}
\usepackage{breqn}
\usepackage{latexsym}
\usepackage{accents}
\usepackage{eufrak}
\usepackage{psfrag}
\usepackage{color}
\usepackage{times,color}
\usepackage[usenames,dvipsnames,svgnames,table]{xcolor}
\usepackage{pgf}
\usepackage{tikz}
\usetikzlibrary{arrows, automata, positioning, calc}
\usepackage{cancel} 
\usepackage{caption}
\usepackage{subcaption}
\usepackage{arydshln}
\usepackage{rotating}
\usepackage[T1]{fontenc}
\usepackage{anyfontsize}
\usepackage[hyphens]{url}
\usepackage[bookmarks=false]{hyperref}

\newcommand{\myLambda}{\begin{sideways}%
     \begin{sideways}$\mathrm{V}$\end{sideways}\end{sideways}}

\usepackage{algorithm,algpseudocode}

\makeatletter
\newcommand{\removelatexerror}{\let\@latex@error\@gobble}
\makeatletter
\newcommand{\proofpart}[2]{%
	\par
	\addvspace{\medskipamount}%
	\noindent\emph{Part #1: #2}\par\nobreak
	\addvspace{\smallskipamount}%
	\@afterheading
}
\makeatother

\makeatletter
\newcommand*{\transpose}{%
  {\mathpalette\@transpose{}}%
}
\newcommand*{\@transpose}[2]{%
  \raisebox{\depth}{$\m@th#1\intercal$}%
}
\makeatother

\renewcommand{\mathsf}[1]{#1}

\theoremstyle{definition}

\IEEEoverridecommandlockouts

\begin{document}

\newcommand{\SB}[3]{
\sum_{#2 \in #1}\biggl|\overline{X}_{#2}\biggr| #3
\biggl|\bigcap_{#2 \notin #1}\overline{X}_{#2}\biggr|
}

\newcommand{\Mod}[1]{\ (\textup{mod}\ #1)}

\newcommand{\overbar}[1]{\mkern 0mu\overline{\mkern-0mu#1\mkern-8.5mu}\mkern 6mu}

\makeatletter
\newcommand*\nss[3]{%
  \begingroup
  \setbox0\hbox{$\m@th\scriptstyle\cramped{#2}$}%
  \setbox2\hbox{$\m@th\scriptstyle#3$}%
  \dimen@=\fontdimen8\textfont3
  \multiply\dimen@ by 4             
  \advance \dimen@ by \ht0
  \advance \dimen@ by -\fontdimen17\textfont2
  \@tempdima=\fontdimen5\textfont2  
  \multiply\@tempdima by 4
  \divide  \@tempdima by 5          
  \ifdim\dimen@<\@tempdima
    \ht0=0pt                        
    \@tempdima=\fontdimen5\textfont2
    \divide\@tempdima by 4          
    \advance \dimen@ by -\@tempdima 
    \ifdim\dimen@>0pt
      \@tempdima=\dp2
      \advance\@tempdima by \dimen@
      \dp2=\@tempdima
    \fi
  \fi
  #1_{\box0}^{\box2}%
  \endgroup
  }
\makeatother

\makeatletter
\renewenvironment{proof}[1][\proofname]{\par
  \pushQED{\qed}%
  \normalfont \topsep6\p@\@plus6\p@\relax
  \trivlist
  \item[\hskip\labelsep
        \itshape
    #1\@addpunct{:}]\ignorespaces
}{%
  \popQED\endtrivlist\@endpefalse
}
\makeatother

\makeatletter
\newsavebox\myboxA
\newsavebox\myboxB
\newlength\mylenA

\newcommand*\xoverline[2][0.75]{%
    \sbox{\myboxA}{$\m@th#2$}%
    \setbox\myboxB\null
    \ht\myboxB=\ht\myboxA%
    \dp\myboxB=\dp\myboxA%
    \wd\myboxB=#1\wd\myboxA
    \sbox\myboxB{$\m@th\overline{\copy\myboxB}$}
    \setlength\mylenA{\the\wd\myboxA}
    \addtolength\mylenA{-\the\wd\myboxB}%
    \ifdim\wd\myboxB<\wd\myboxA%
       \rlap{\hskip 0.5\mylenA\usebox\myboxB}{\usebox\myboxA}%
    \else
        \hskip -0.5\mylenA\rlap{\usebox\myboxA}{\hskip 0.5\mylenA\usebox\myboxB}%
    \fi}
\makeatother

\xpatchcmd{\proof}{\hskip\labelsep}{\hskip3.75\labelsep}{}{}

\pagestyle{plain}

\title{\fontsize{21}{28}\selectfont Sparse Random Khatri-Rao Product Codes for\\ Distributed Matrix Multiplication}

\author{Ruowan Ji, Anoosheh Heidarzadeh, and Krishna R. Narayanan\thanks{The authors are with the Department of Electrical and Computer Engineering, Texas A\&M University, College Station, TX 77843 USA (E-mail: \{jiruowan, anoosheh, krn\}@tamu.edu).}}

\maketitle 

\thispagestyle{plain}

\begin{abstract}
We introduce two generalizations to the paradigm of using Random Khatri-Rao Product (RKRP) codes for distributed matrix multiplication. 
We first introduce a class of codes called Sparse Random Khatri-Rao Product (SRKRP) codes which have sparse generator matrices. 
SRKRP codes result in lower encoding, computation and communication costs than RKRP codes when the input matrices are sparse, while they 
exhibit similar numerical stability to other state of the art schemes.
We empirically study the relationship between the probability of the generator matrix (restricted to the set of non-stragglers) of a randomly chosen SRKRP code being rank deficient and various parameters of the coding scheme including the degree of sparsity of the generator matrix and the number of non-stragglers. 
Secondly, we show that if the master node can perform a very small number of
matrix product computations in addition to the computations performed by the workers, the failure probability can be substantially improved.
\end{abstract}

\section{Introduction}
Many machine learning applications require multiplication of two large matrices with real-valued entries. 
Such large-scale matrix multiplications cannot be simply performed on a single machine, and a natural solution is to parallelize the computation using the master-worker paradigm on distributed computing platforms. 
In classical distributed matrix multiplication schemes, the master splits each of the two input matrices into smaller blocks (submatrices), and requests each worker to compute and return the product of a pair of blocks---each belonging to one of the two input matrices. 
Upon receiving the computation results of all workers, the master recovers the product of the two input matrices. 
However, such systems are prone to stragglers (i.e., those workers that do not return their results as quickly as the rest of the workers) because  
the master must wait for all workers---including the stragglers---to finish their computations and return their results~\cite{LLPPR2018}.

A promising approach to mitigate the effect of stragglers is to incorporate redundancy in the computations of the workers---using coding techniques---so that the master can recover the required product from the results of a subset of workers, instead of waiting for the results of all workers~\cite{RDT2020}. 
Inspired by the work of Lee \emph{et al.}~\cite{LLPPR2018}, several coding-based distributed matrix multiplication schemes have been recently proposed~\cite{YMAA2017,DBJLG2018,DFHJCG2020,YA2020,FC2021,DRV2021,RT2022,LSR2017,BLOR2018,WLS2018,PHN2021,SHN2019}. 
(Several different variations of the coded distributed matrix multiplication problem---not closely related to our work---have also been studied in the literature, see, e.g.,~\cite{DR2022,KFD2021,JPHN2021,HYL2020,DTR2018,CT2019No2,GWCR2018,JNMA2021,JNMA2022,SMA2021,JDCC2021,KFD2018,DOERK2019,ASK2020,CT2019,JJ2021,DOERHK2021,HGVG2022,KKES2021,KYL2019,CJWJ2020,YL2019,KL2019,DCG2019,YLRKSA2019,JJ2021No2,JJ2020}.) 
These schemes provide different trade-offs between several performance metrics including
(i) recovery threshold, i.e., the minimum number of non-straggling workers required for successful recovery,
(ii) communication cost, i.e., the average amount of information that needs to be transferred from the master to a worker, 
(iii) computation load, i.e., the average number of arithmetic operations performed by a worker, 
(iv) computational complexity of encoding and decoding processes, and 
(v) numerical stability in the presence of round-off and truncation errors.

Most of the existing codes for distributed matrix multiplication provide deterministic guarantees on the recovery threshold, i.e., the master can decode from the results of \emph{any} subset of workers of size no less than a certain threshold. 
Examples of such coding schemes are Polynomial codes \cite{YMAA2017} and MatDot codes~\cite{DFHJCG2020}. 
While these codes have excellent performance in terms of recovery threshold, they are highly numerically unstable when the operations are performed over the real field. 
Motivated by this, several numerically-stable coding schemes with deterministic guarantees were recently proposed in~\cite{FC2021,DRV2021,RT2022}. 
A comprehensive comparison of codes with deterministic guarantees, which we collectively refer to as \emph{deterministic codes}, can be found in~\cite{PHN2021}. 

Aside from deterministic codes are the coding schemes that provide probabilistic guarantees on the recovery threshold, i.e., the master can decode from the results of \emph{a randomly chosen} subset of workers of size no less than a certain threshold, \emph{with high probability}. 
Examples of such codes include Sparse codes~\cite{WLS2018}, Factored Luby-Transform (FLT) codes and Factored Raptor (FRT) codes~\cite{PHN2021}, and Random Khatri-Rao Product (RKRP) codes~\cite{SHN2019}. 
All of these codes are highly numerically stable. 
Sparse codes and FLT/FRT codes achieve optimal recovery threshold asymptotically (with probability approaching $1$) as the number of workers grows unbounded, whereas RKRP codes have optimal recovery threshold with probability $1$.
RKRP codes have a dense generator matrix, whereas Sparse codes and FLT/FRT codes have sparse generator matrices. 
As a result, the encoding/decoding complexity and the computation load of these codes can be substantially lower than those of RKRP codes and deterministic codes, particularly when the input matrices are sparse~\cite{PHN2021}. 
The main difference between Sparse codes and FLT/FRT codes is that 
the communication cost of Sparse codes is substantially higher than that of FLT/FRT codes (or even RKRP codes and deterministic codes), whereas the communication cost of FLT/FRT codes can be much lower than that of RKRP codes and deterministic codes, particularly when the input matrices are sparse~\cite{PHN2021}.



In this work, we introduce a new coding scheme, referred to as \emph{Sparse Random Khatri-Rao Product (SRKRP) codes}, which is a generalization of RKRP codes.
An SRKRP code can have a very sparse generator matrix---similar to FLT/FRT codes. 
As a result, when the input matrices are sparse, 
the encoding complexity, the communication cost, and the computation cost of SRKRP codes can be much lower than those of the original RKRP codes.
The decoding complexity of SRKRP codes is, however, comparable to that of the original RKRP codes, and 
higher than that of FLT/FRT codes. 
The numerical stability of SRKRP codes is also comparable to that of RKRP codes and FLT/FRT codes.

When compared to FLT/FRT codes, SRKRP codes---with generator matrices of the same size and the same degree of sparsity---have a substantially lower failure probability, even when the number of workers is in the order of tens or hundreds. 
While a theoretical analysis of the failure probability of SRKRP codes remains unknown in general, our simulations show that these codes can have a very low failure probability, 
even when the generator matrix of the code is much sparser than that of the original RKRP codes. 
In addition, our simulations show that a few extra computations (as little as one) performed locally at the master---in parallel to those computations performed by the workers---can substantially reduce the failure probability of SRKRP codes. 
To the best of our knowledge, this work is the first in the literature on coded distributed matrix multiplication to study the role of such extra computations.

\section{Problem Setup}\label{sec:PS}
We use bold-face capital (lowercase) letters for matrices (vectors). 
We denote the entry $(a,b)$ of matrix $\mathbf{M}$ by $(\mathbf{M})_{a,b}$. 
For any integers $1<i<j$, we denote $\{i,i+1,\dots,j\}$ by $[i:j]$, and for any integer $i\geq 1$, denote $\{1,\dots,i\}$ by $[i]$.

Consider a distributed master-worker framework in which the master node has two input matrices $\mathbf{A}\in \mathbb{R}^{r\times s}$ and $\mathbf{B}\in \mathbb{R}^{r\times t}$, and wishes to compute the matrix  $\mathbf{C}:=\mathbf{A}^{\mathsf{T}}\mathbf{B}$ using the help of $N$ worker nodes. 
To do so, suppose that the master node splits the input matrix $\mathbf{A}$ column-wise into $m$ submatrices $\mathbf{A}_1,\dots,\mathbf{A}_m\in \mathbb{R}^{r\times \frac{s}{m}}$, 
and splits the input matrix $\mathbf{B}$ column-wise into $n$ submatrices $\mathbf{B}_1,\dots,\mathbf{B}_n\in \mathbb{R}^{r\times \frac{t}{n}}$, 
where $m,n$ are two arbitrary integers such that $mn\leq N$. 
Note that the matrix $\mathbf{C} = \mathbf{A}^{\mathsf{T}}\mathbf{B} = [\mathbf{A}^{\mathsf{T}}_i\mathbf{B}_j]_{i\in [m],j\in [n]}$. 
Thus, in order to compute $\mathbf{C}$, the master node uses the help of the worker nodes to compute the $K:=mn$ smaller matrix multiplications $\{\mathbf{A}^{\mathsf{T}}_i \mathbf{B}_j\}_{i\in [m], j\in [n]}$. 

Suppose that the computations performed by a randomly chosen subset of $S$ worker nodes---whose identities are initially unknown at the master node---are subject to erasure. 
Such worker nodes are referred to as \emph{stragglers} in the literature on distributed computing. 
Due to the existence of stragglers, the master node cannot simply request the worker nodes to compute the smaller matrix multiplications $\mathbf{A}^{\mathsf{T}}_i \mathbf{B}_j$. 
Instead, 
the master node first encodes the $m$ submatrices $\mathbf{A}_1,\dots,\mathbf{A}_m$ and the $n$ submatrices $\mathbf{B}_1,\dots,\mathbf{B}_n$ into $N$ coded submatrices $\widetilde{\mathbf{A}}_1,\dots,\widetilde{\mathbf{A}}_{N}\in \mathbb{R}^{r\times \frac{s}{m}}$ and $N$ coded submatrices $\widetilde{\mathbf{B}}_1,\dots,\widetilde{\mathbf{B}}_{N}\in \mathbb{R}^{r\times \frac{t}{n}}$, respectively. 
Then, for each $l\in [N]$, the master node sends $\widetilde{\mathbf{A}}_l$ and $\widetilde{\mathbf{B}}_l$ to the worker node $l$, and requests the worker node $l$ to compute $\widetilde{\mathbf{A}}^{\mathsf{T}}_l \widetilde{\mathbf{B}}_l$ and send the result back to the master node. 
For each $l\in [N]$, let 
\begin{equation}\label{eq:n1}
\widetilde{\mathbf{A}}_l:=\sum_{i=1}^{m} p_{l,i}\mathbf{A}_{i}, \quad \text{and} \quad  \widetilde{\mathbf{B}}_l:=\sum_{j=1}^{n} q_{l,j}\mathbf{B}_{j},
\end{equation} 
where $\mathbf{p}_l:=[p_{l,1},\dots,p_{l,m}]$ and $\mathbf{q}_l:=[q_{l,1},\dots,q_{l,n}]$ are two row-vectors with real entries representing the coding coefficients pertaining to $\widetilde{\mathbf{A}}_l$ and $\widetilde{\mathbf{B}}_l$, respectively.  

In addition to the help from the worker nodes, in this work we assume that the master node can also perform some computations locally. 
To be more specific, we consider the case in which the master node can perform $R$ extra computations $\widetilde{\mathbf{A}}^{\mathsf{T}}_{N+1}\widetilde{\mathbf{B}}_{N+1},\dots,\widetilde{\mathbf{A}}^{\mathsf{T}}_{N+R}\widetilde{\mathbf{B}}_{N+R}$ in parallel. 
For each $l\in [N+1:N+R]$, the coded submatrices $\widetilde{\mathbf{A}}_{l}$ and $\widetilde{\mathbf{B}}_{l}$ are constructed similarly as in~\eqref{eq:n1}, and the coding vectors $\mathbf{p}_l$ and $\mathbf{q}_l$ corresponding to $\widetilde{\mathbf{A}}_{l}$ and $\widetilde{\mathbf{B}}_{l}$ are defined as before.    
Note that the extra computations performed by the master node are not subject to erasures. 
That said, these computations are designed in advance---without the knowledge of the configuration of stragglers, and are performed in parallel to those computations performed by the worker nodes. 

The goal is to design an encoding scheme, i.e., a (potentially randomized) algorithm for generating the coding vectors $\mathbf{p}_l$'s and $\mathbf{q}_l$'s 
such that the master node can successfully recover $\{\mathbf{A}^{\mathsf{T}}_i \mathbf{B}_j\}_{i\in [m], j\in [n]}$ by decoding the results of the $N-S$ computations performed by the non-straggling worker nodes and the results of the $R$ extra computations performed by the master node.

For each $l\in [N+R]$, let $\widetilde{\mathbf{C}}_{l}:=\widetilde{\mathbf{A}}^{\mathsf{T}}_l \widetilde{\mathbf{B}}_l$.
The results received by the master node and those computed locally at the master node can be written in matrix form as follows: 
\begin{equation}\label{eq:n2}
\begin{bmatrix}
\widetilde{\mathbf{C}}_{l_1}\\ 
\vdots\\ 
\widetilde{\mathbf{C}}_{l_{N-S}}\\[0.125cm] \hdashline[2pt/2pt]\\[-0.3cm] \widetilde{\mathbf{C}}_{N+1}\\ \vdots \\ \widetilde{\mathbf{C}}_{N+R}
\end{bmatrix} =
\begin{bmatrix}
\mathbf{p}_{l_1}\otimes \mathbf{q}_{l_1}\\ 
\vdots\\ \mathbf{p}_{l_{N-S}}\otimes \mathbf{q}_{l_{N-S}}\\[0.125cm] \hdashline[2pt/2pt]\\[-0.3cm] \mathbf{p}_{N+1}\otimes \mathbf{q}_{N+1}\\ \vdots \\ \mathbf{p}_{N+R}\otimes \mathbf{q}_{N+R}
\end{bmatrix}
\begin{bmatrix}
\mathbf{A}^{\mathsf{T}}_1\mathbf{B}_1\\
\vdots\\
\mathbf{A}^{\mathsf{T}}_1\mathbf{B}_n\\
\vdots\\
\mathbf{A}^{\mathsf{T}}_m\mathbf{B}_1\\
\vdots\\
\mathbf{A}^{\mathsf{T}}_m\mathbf{B}_n
\end{bmatrix},
\end{equation} 
where $l_1,l_2,\dots,l_{N-S}\in [N]$ represent the indices of the $N-S$ non-straggling worker nodes, and $\mathbf{p}\otimes\mathbf{q}$ represents the Kronecker product of the row-vectors $\mathbf{p}$ and $\mathbf{q}$, i.e., \[\mathbf{p}\otimes \mathbf{q} = [p_1q_1,p_1q_2,\dots,p_1q_n,\dots,p_mq_1,p_mq_2,\dots,p_mq_n],\] where $\mathbf{p} = [p_1,\dots,p_m]$ and $\mathbf{q}=[q_1,\dots,q_n]$. 
One can easily observe that the decoding is successful if and only if the coefficient matrix in the system of linear equations~\eqref{eq:n2} is full-rank. 
When the rank is full, the master node solves the system of linear equations in~\eqref{eq:n2} and obtains an estimate of $\mathbf{C}_{i,j}:=\mathbf{A}^{\mathsf{T}}_i\mathbf{B}_j$, denoted by $\widehat{\mathbf{C}}_{i,j}$, for each $i\in [m]$ and each $j\in [n]$. 
(Since the operations are performed over $\mathbb{R}$, the computations are prone to numerical errors, and 
hence, $\mathbf{C}_{i,j}$ and $\widehat{\mathbf{C}}_{i,j}$ may not necessarily be equal.)  
An estimate of $\mathbf{C}=\mathbf{A}^{\mathsf{T}}\mathbf{B}$ is then obtained by $\widehat{\mathbf{C}}:=[\widehat{\mathbf{C}}_{i,j}]_{i\in [m], j\in [n]}$. 


\section{Proposed Coding Scheme}\label{sec:PCS}
We build upon RKRP codes of~\cite{SHN2019}, and propose a generalization of these codes, referred to as \emph{Sparse RKRP (SRKRP) codes}, which can have a sparse generator matrix. 

\subsection{Encoding}\label{subsec:Enc}
Let $\mathrm{U}(x) :=\sum_{k=1}^{m} \mathrm{U}_{k} x^{k}$ and $\mathrm{V}(x) :=\sum_{k=1}^{n} \mathrm{V}_{k} x^{k}$ be the polynomial representation of two weight distributions, i.e., 
$0\leq \mathrm{U}_k\leq 1$ for all $k\in [m]$, $0\leq \mathrm{V}_k\leq 1$ for all $k\in [n]$, and ${\sum_{k=1}^{m}\mathrm{U}_k = \sum_{k=1}^{n}\mathrm{V}_k = 1}$. 
Similarly, let $\mathrm{U}^{*}(x):=\sum_{k=1}^{m} \mathrm{U}^{*}_{k} x^{k}$ and $\mathrm{V}^{*}(x):=\sum_{k=1}^{n} \mathrm{V}^{*}_{k} x^{k}$ be the polynomial representation of two weight distributions. 

Let $X$ be an arbitrary random variable such that the CDF of $X$ is absolutely continuous with respect to the Lebesgue measure, e.g., the uniform random variable $X\sim \mathcal{U}(0,1)$. 

In an SRKRP code, the coding vectors $\mathbf{p}_l = [p_{l,1},\dots,p_{l,m}]$ and $\mathbf{q}_l = [q_{l,1},\dots,q_{l,n}]$ for each $l\in [N]$ are constructed as follows: 
\begin{enumerate}
    \item Randomly choose a weight $u_l$ and a weight $v_l$ by sampling from the weight distribution $\mathrm{U}(x)$ and the weight distribution $\mathrm{V}(x)$, respectively, where the probability of $u_l=k$ is $\mathrm{U}_k$ for each $k\in [m]$, and the probability of $v_l=k$ is $\mathrm{V}_k$ for each $k\in [n]$. 
    \item Randomly choose a subset of $[m]$ of size $u_l$, say, ${\mathcal{S}_l}$, and randomly choose a subset of $[n]$ of size $v$, say, ${\mathcal{T}_l}$. 
    \item Let $\{p_{l,i}: i\in \mathcal{S}_l\}$ 
    and $\{q_{l,j}: j\in \mathcal{T}_l\}$ be independently generated realizations of random variable $X$. 
    Also, let ${p_{l,i} = 0}$ for all ${i\not\in \mathcal{S}_l}$, and let ${q_{l,j} = 0}$ for all ${j\not\in \mathcal{T}_l}$.
\end{enumerate}

The coding vectors $\mathbf{p}_{l}$ and $\mathbf{q}_{l}$ for each $l\in [N+1:N+R]$ are also constructed similarly as above except that in this case the weight distributions $\mathrm{U}(x)$ and $\mathrm{V}(x)$ are replaced by the weight distributions $\mathrm{U}^{*}(x)$ and $\mathrm{V}^{*}(x)$, respectively. 

Let $l_1,\dots,l_{N-S}\in [N]$ be the indices of the $N-S$ non-straggling worker nodes. 
Let $\mathbf{P}$ and $\mathbf{Q}$ be two matrices defined as 
$\mathbf{P}=[\mathbf{p}_{l_1}^{\mathsf{T}},\dots,\mathbf{p}_{l_{N-S}}^{\mathsf{T}},\mathbf{p}_{N+1}^{\mathsf{T}},\dots,\mathbf{p}_{N+R}^{\mathsf{T}}]^{\mathsf{T}}$ 
and 
$\mathbf{Q}=[\mathbf{q}_{l_1}^{\mathsf{T}},\dots,\mathbf{q}_{l_{N-S}}^{\mathsf{T}},\mathbf{q}_{N+1}^{\mathsf{T}},\dots,\mathbf{q}_{N+R}^{\mathsf{T}}]^{\mathsf{T}}$. 
Note that the size of $\mathbf{P}$ is $(N-S+R)\times m$, and the size of $\mathbf{Q}$ is $(N-S+R)\times n$. 
Each of the first $N-S$ rows of $\mathbf{P}$ contains $u_{\text{avg}} =\sum_{k=1}^{m} \mathrm{U}_k k$ nonzero entries on average, and each of the last $R$ rows of $\mathbf{P}$ contains $u^{*}_{\text{avg}} =\sum_{k=1}^{m} \mathrm{U}^{*}_k k$ nonzero entries on average. 
Similarly, each of the first $N-S$ rows of $\mathbf{Q}$ contains $v_{\text{avg}} =\sum_{k=1}^{n} \mathrm{V}_k k$ nonzero entries on average, and each of the last $R$ rows of $\mathbf{Q}$ contains $v^{*}_{\text{avg}} = \sum_{k=1}^{n} \mathrm{V}^{*}_k k$ nonzero entries on average.
Let $\mathbf{G} := \mathbf{P}\odot\mathbf{Q}$ be the row-wise Khatri-Rao product of the matrices $\mathbf{P}$ and $\mathbf{Q}$, i.e., 
\begin{equation}\label{eq:n3}
\mathbf{G} = 
\begin{bmatrix}
\mathbf{p}_{l_1}\otimes \mathbf{q}_{l_1}\\
\vdots\\
\mathbf{p}_{l_{N-S}}\otimes \mathbf{q}_{l_{N-S}}\\
\mathbf{p}_{N+1}\otimes \mathbf{q}_{N+1}\\
\vdots\\
\mathbf{p}_{N+R}\otimes \mathbf{q}_{N+R}
\end{bmatrix}.
\end{equation} 
It is easy to verify that each of the first $N-S$ rows of $\mathbf{G}$ contains $w_{\text{avg}}=u_{\text{avg}}v_{\text{avg}}$ nonzero entries on average, and each of the last $R$ rows of $\mathbf{G}$ contains $w^{*}_{\text{avg}} =u^{*}_{\text{avg}}v^{*}_{\text{avg}}$ nonzero entries on average. 
Recall that in the original RKRP codes~\cite{SHN2019}, there are no weight distributions $\mathrm{U}^{*}(x)$ and $\mathrm{V}^{*}(x)$ since $R=0$, and the weight distributions $\mathrm{U}(x) = x^m$ and $\mathrm{V}(x) = x^n$. 
Note that in this case, ${u_{\text{avg}} = m}$ and ${v_{\text{avg}} = n}$. 
This implies that the coding vectors in the original RKRP codes are dense. 
Taking $\mathrm{U}(x)$ and $\mathrm{V}(x)$ to be weight distributions with $u_{\text{avg}}<m$ and $v_{\text{avg}}<n$, SRKRP codes can take advantage of sparser coding vectors when compared to the original RKRP codes. 

\subsection{Decoding}\label{subsec:Dec}
Note that the matrix $\mathbf{G}$ defined as in~\eqref{eq:n3} is the coefficient matrix in the system of linear equations~\eqref{eq:n2}.
Rewriting~\eqref{eq:n2}, for each $a\in [\frac{s}{m}]$ and each $b\in [\frac{t}{n}]$, we have 
\begin{equation}\label{eq:n4}
\underbrace{
\begin{bmatrix}
(\widetilde{\mathbf{C}}_{l_1})_{a,b}\\ 
\vdots\\
(\widetilde{\mathbf{C}}_{l_{N-S}})_{a,b}\\
(\widetilde{\mathbf{C}}_{N+1})_{a,b}\\
\vdots\\
(\widetilde{\mathbf{C}}_{N+R})_{a,b}
\end{bmatrix}
}_{\mathbf{y}^{(a,b)}}
= \mathbf{G} 
\underbrace{
\begin{bmatrix}
(\mathbf{C}_{1,1})_{a,b}\\
\vdots\\
(\mathbf{C}_{1,n})_{a,b}\\
\vdots\\
(\mathbf{C}_{m,1})_{a,b}\\
\vdots\\
(\mathbf{C}_{m,n})_{a,b}\\
\end{bmatrix}
}_{\mathbf{z}^{(a,b)}},
\end{equation} where 
$\widetilde{\mathbf{C}}_l$'s and $\mathbf{C}_{i,j}$'s are as defined in Section~\ref{sec:PS}.  
Note that $\mathbf{y}^{(a,b)}$ and $\mathbf{z}^{(a,b)}$ defined in~\eqref{eq:n4} are two column-vectors with real entries, each of length $K$; and all $K$ coordinates of $\mathbf{y}^{(a,b)}$ are known by the master node, whereas the $K$ coordinates of $\mathbf{z}^{(a,b)}$ are unknown at the master node. 
Given that the matrix $\mathbf{G}$ is full-rank, 
the master node solves the system of linear equations~\eqref{eq:n4}, and 
obtains an estimate $\widehat{\mathbf{z}}^{(a,b)}$ of $\mathbf{z}^{(a,b)}$. 
(In the absence of numerical errors, $\widehat{\mathbf{z}}^{(a,b)} = \mathbf{z}^{(a,b)}$.)
Upon computing $\widehat{\mathbf{z}}^{(a,b)}$ for all $a\in [\frac{s}{m}]$ and all $b\in [\frac{t}{n}]$, the master node obtains an estimate $\widehat{\mathbf{C}} = [\widehat{\mathbf{C}}_{i,j}]_{i\in [m], j\in [n]}$ of $\mathbf{C}$, where $(\widehat{\mathbf{C}}_{i,j})_{a,b}$ 
is the ${((i-1)n+j)}$th coordinate of $\widehat{\mathbf{z}}^{(a,b)}$.   



\section{Performance Analysis}\label{sec:PA}
To measure the performance of SRKRP codes, we consider the following metrics: 
(i)~failure probability, 
(ii)~computation load per worker, 
(iii)~communication cost per worker, 
(iv)~encoding and decoding complexity, and 
(v)~numerical stability.  

\subsection{Failure probability}\label{subsec:FP}
Since the encoding scheme of SRKRP codes is randomized and the configuration of stragglers is assumed to be random, the decoding may or may not be successful for a given realization of the coding vectors and a given configuration of the stragglers. 
As a result, we consider the failure probability---defined as the probability that the decoding fails for a randomly generated code realization and a randomly chosen configuration of stragglers---as a metric to measure the performance of SRKRP codes. 

Thinking of $\mathbf{P}$ and $\mathbf{Q}$ as two random matrices, it can be seen that the structure of the matrix $\mathbf{G} = \mathbf{P}\odot \mathbf{Q}$ is random, and hence, the rank of $\mathbf{G}$ is a random variable.
Thus, the failure probability is equal to the probability that a randomly generated matrix $\mathbf{G}$ is not full-rank. 

\vspace{0.125cm}
{\bf No extra computation ($\boldsymbol{R}\boldsymbol{=}\boldsymbol{0}$):}
As was shown in~\cite{SHN2019}, for the case of $R=0$, the matrix $\mathbf{G}$ is full-rank with probability $1$ when $\mathrm{U}(x)=x^m$ and $\mathrm{V}(x)=x^n$. 
Also, when $\mathrm{U}(x)=x$ and $\mathrm{V}(x)=x$, 
it is easy to show that if $N-S=K$, the matrix $\mathbf{G}$ is full-rank with probability ${K!}/{K^K}$, which converges to $0$ as $K$ grows unbounded. 
To the best of our knowledge, the full-rank probability of the matrix $\mathbf{G}$ is not known for any other $\mathrm{U}(x)$ and $\mathrm{V}(x)$, and hence,
a theoretical analysis of the failure probability of SRKRP codes remains unknown in general. 
Notwithstanding, our simulation results in Section~\ref{sec:SIM} reveal several interesting properties of these codes for the case of $R=0$:
\begin{enumerate}
    \item The failure probability depends mainly on the average weights ${u_{\text{avg}}}$ and ${v_{\text{avg}}}$, and does not change for different pairs of $\mathrm{U}(x)$ and $\mathrm{V}(x)$ which yield the same average weights ${u_{\text{avg}}}$ and ${v_{\text{avg}}}$, respectively.
    \item For a fixed overall average weight $w_{\text{avg}}$, the closer are the average weights $u_{\text{avg}}$ and $v_{\text{avg}}$ to $\sqrt{w_{\text{avg}}}$, the smaller is the failure probability. 
    \item For 
    ${w_{\text{avg}}>\log K}$, 
    the failure probability is close to the probability that the matrix $\mathbf{G}$ has an all-zero column, and the latter probability is close to the probability that a random matrix $\mathbf{M}$ of the same size as the matrix $\mathbf{G}$ has an all-zero column, where the entries of the matrix $\mathbf{M}$ are realizations of i.i.d.~Bernoulli random variables with success probability ${w_{\text{avg}}}/{K}$ (i.e., each entry of the matrix $\mathbf{M}$ is $1$ with probability ${w_{\text{avg}}}/{K}$, independent of the other entries).
\end{enumerate}

Observations (1) and (2) suggest that without loss of generality, we can consider the same weight distribution for both $\mathrm{U}(x)$ and $\mathrm{V}(x)$, i.e., $\mathrm{U}(x)=\mathrm{V}(x)=\myLambda(x)$ for some weight distribution $\myLambda(x)$, and 
we can restrict our attention to weight distributions $\myLambda(x)$ of simplest form that yield the overall average weight $w_{\text{avg}}$, 
i.e., 
${\myLambda(x)=x^{\sqrt{w_{\text{avg}}}}}$ or ${\myLambda(x)=\lambda x^{\lfloor\sqrt{w_{\text{avg}}}\rfloor}+(1-\lambda)x^{\lceil\sqrt{w_{\text{avg}}}\rceil}}$ for ${\lambda = (\lceil \sqrt{w_{\text{avg}}}\rceil - \sqrt{w_{\text{avg}}})/(\lceil \sqrt{w_{\text{avg}}}\rceil - \lfloor \sqrt{w_{\text{avg}}} \rfloor)}$ when $w_{\text{avg}}$ is a perfect square or not a perfect square, respectively.  

Observation (3) suggests that the failure probability of an SRKRP code with $R=0$ can be closely approximated by ${1-(1-(1-w_{\text{avg}}/K)^{N-S})^{K}}$. 
This is because a column of a random matrix of size ${(N-S)\times K}$---whose entries are realizations of i.i.d.~Bernoulli random variables with success probability $w_{\text{avg}}/K$---is all-zero with probability ${(1-w_{\text{avg}}/K)^{N-S}}$, and hence, all $K$ columns of such a matrix are nonzero with probability ${(1-(1-w_{\text{avg}}/K)^{N-S})^{K}}$. 

\vspace{0.125cm}
{\bf Leveraging extra computations ($\boldsymbol{R}\boldsymbol{\geq}\boldsymbol{1}$):}
The above observation implies that when $R=0$ and $w_{\text{avg}}<K$, an SRKRP code 
fails with a nonzero probability.
When $K$ grows unbounded and ${N-S-K}$ remains constant, if ${w_{\text{avg}}>\log K}$, the failure probability vanishes. 
However, when $K$ is finite and $N-S-K$ is small, 
the failure probability may not be as small as required, even for arbitrarily large $w_{\text{avg}}<K$. 
To alleviate this drawback, we propose to leverage $R$ extra computations performed by the master node.

Extending the result of~\cite{SHN2019} to the cases with $R\geq 1$, it is immediate that the matrix $\mathbf{G}$ is full-rank with probability~$1$ when $\mathrm{U}(x)=\mathrm{U}^{*}(x)=x^m$ and $\mathrm{V}(x)=\mathrm{V}^{*}(x)=x^n$.
However, the probability of the matrix $\mathbf{G}$ being full-rank remains unknown for any $\mathrm{U}(x)$ and $\mathrm{V}(x)$ with ${u_{\text{avg}}<m}$ and ${v_{\text{avg}}<n}$, even when $\mathrm{U}^{*}(x)=x^m$ and $\mathrm{V}^{*}(x)=x^n$.  
Intuitively, for any $u_{\text{avg}}$ and $v_{\text{avg}}$ and any $R\geq 1$, we expect that the larger are the average weights $u^{*}_{\text{avg}}$ and $v^{*}_{\text{avg}}$, the larger is the probability that the matrix $\mathbf{G}$ is full-rank. 
Our simulation results in Section~\ref{sec:SIM} are consistent with this intuition. 
Moreover, the results of our simulations show that a few extra computations (i.e., ${R\in \{1,2\}}$) with sufficiently large $w^{*}_{\text{avg}}$ can significantly reduce the failure probability, 
even when $w_{\text{avg}}$ is as small as $\log K$.



\subsection{Computation load per worker}\label{subsec:CompL} 
Another performance metric that we consider is the average computational complexity of the matrix multiplication performed by a worker node. 
Let ${nnz}(\mathbf{A})$ and ${nnz}(\mathbf{B})$ denote the number of nonzero entries in $\mathbf{A}$ and $\mathbf{B}$, respectively. 
For the ease of exposition, assume that the positions of the nonzero entries of $\mathbf{A}$ and $\mathbf{B}$ are randomly chosen. 
Note that ${nnz}(\mathbf{A}_i)$ is $\mathcal{O}(\frac{nnz(\mathbf{A})}{m})$ for all $i\in [m]$, and ${nnz}(\mathbf{B}_j)$ is $\mathcal{O}(\frac{nnz(\mathbf{B})}{n})$ for all $j\in [n]$. 
For a given $l\in [N]$, let $u_l$ and $v_l$ be the number of nonzero coordinates in the coding vectors $\mathbf{p}_l$ and $\mathbf{q}_l$, respectively. 
Then, $nnz(\widetilde{\mathbf{A}}_l)$ and $nnz(\widetilde{\mathbf{B}}_l)$ are $\mathcal{O}(u_l \frac{nnz(\mathbf{A})}{m})$ and  $\mathcal{O}(v_l \frac{nnz(\mathbf{B})}{n})$, respectively.
This further implies that the complexity of computing  $\widetilde{\mathbf{A}}^{\mathsf{T}}_l \widetilde{\mathbf{B}}_l$ is $\mathcal{O}(\min\{u_l t \frac{nnz(\mathbf{A})}{mn},v_l s \frac{nnz(\mathbf{B})}{mn}\})$. 
Thus, the computation load per worker is given by  \[\mathcal{O}\left(\min\left\{u_{\text{avg}} t \frac{nnz(\mathbf{A})}{mn},v_{\text{avg}} s \frac{nnz(\mathbf{B})}{mn}\right\}\right).\] 

\subsection{Communication cost per worker}
The communication cost per worker is defined as the average amount of data that needs to be transferred from the master node to a worker node. 
For a given $l\in [N]$, the master node sends $\widetilde{\mathbf{A}}_l$ and $\widetilde{\mathbf{B}}_l$ to the worker node $l$. 
Thus, the communication cost per worker node $l$ is $nnz(\widetilde{\mathbf{A}}_l)+nnz(\widetilde{\mathbf{B}}_l)$. 
Since $nnz(\widetilde{\mathbf{A}}_l)$ and $nnz(\widetilde{\mathbf{B}}_l)$ are $\mathcal{O}(u_l \frac{nnz(\mathbf{A})}{m})$ and  $\mathcal{O}(v_l \frac{nnz(\mathbf{B})}{n})$, respectively, the communication cost per worker node $l$ is 
${\mathcal{O}(u_l\frac{nnz(\mathbf{A})}{m}+v_l\frac{nnz(\mathbf{B})}{n})}$. 
Thus, the communication cost per worker is given by
\[\mathcal{O}\left(u_{\text{avg}}\frac{nnz(\mathbf{A})}{m}+v_{\text{avg}}\frac{nnz(\mathbf{B})}{n}\right).\]

\subsection{Encoding and decoding complexity}\label{subsec:EDC} 
The computational complexity of the encoding scheme depends on the sparsity of the input matrices and the sparsity of the coding vectors. 
More specifically, for a given ${l\in [N+R]}$, the computational complexity of encoding $\mathbf{A}_1,\dots,\mathbf{A}_m$ into ${\widetilde{\mathbf{A}}_l}$ is ${\mathcal{O}((2u_l-1)\frac{nnz(\mathbf{A})}{m})}$, and the computational complexity of encoding $\mathbf{B}_1,\dots,\mathbf{B}_n$ into ${\widetilde{\mathbf{B}}_l}$ is ${\mathcal{O}((2v_l-1)\frac{nnz(\mathbf{B})}{n})}$, where $u_l$ and $v_l$ are the number of nonzero coordinates in the coding vectors $\mathbf{p}_l$ and $\mathbf{q}_l$, respectively, and $nnz(\mathbf{A})$ and $nnz(\mathbf{B})$ are the number of nonzero entries in the matrices $\mathbf{A}$ and $\mathbf{B}$, respectively. 
Recall that $u_{\text{avg}}$ and $v_{\text{avg}}$ are the average number of nonzero coordinates over all coding vectors $\mathbf{p}_1,\dots,\mathbf{p}_{N}$ and 
over all coding vectors $\mathbf{q}_1,\dots,\mathbf{q}_{N}$, respectively, and 
${u}^{*}_{\text{avg}}$ and ${v}^{*}_{\text{avg}}$ are the average number of nonzero coordinates over all coding vectors $\mathbf{p}_{N+1},\dots,\mathbf{p}_{N+R}$ and 
over all coding vectors $\mathbf{q}_{N+1},\dots,\mathbf{q}_{N+R}$, respectively. 
Thus, the overall encoding complexity is given by
\[\mathcal{O}\left(N\left((2u_{\text{avg}}-1)\frac{nnz(\mathbf{A})}{m}+(2v_{\text{avg}}-1)\frac{nnz(\mathbf{B})}{n}\right)\right.+\left.R\left((2u^{*}_{\text{avg}}-1)\frac{nnz(\mathbf{A})}{m}+(2v^{*}_{\text{avg}}-1)\frac{nnz(\mathbf{B})}{n}\right)\right).\]


For decoding, the master node needs to solve the system of linear equations in~\eqref{eq:n4}. 
The master node first computes the (pseudo-) inverse of the matrix $\mathbf{G}$, denoted by $\mathbf{G}^{+}$. 
While the matrix $\mathbf{G}$ can be very sparse (depending on the choice of $u_{\text{avg}},v_{\text{avg}},u^{*}_{\text{avg}},v^{*}_{\text{avg}}$), the nonzero entries of the matrix $\mathbf{G}$ are not positioned so that the matrix $\mathbf{G}$ is necessarily well-structured (e.g., banded or block diagonal).
As a result, the complexity of computing the ${K\times (N-S+R)}$ matrix $\mathbf{G}^{+}$ is the same as the complexity of computing the (pseudo-) inverse of a fully dense ${(N-S+R)\times K}$ matrix, i.e., ${\mathcal{O}(K^2(N-S+R))}$. 
For each ${a\in [\frac{s}{m}]}$ and each ${b\in [\frac{t}{n}]}$, the master node then recovers the vector $\mathbf{z}^{(a,b)}$ by multiplying the matrix $\mathbf{G}^{+}$ by the vector $\mathbf{y}^{(a,b)}$.
Recall that for each ${l\in [N]}$, $nnz(\widetilde{\mathbf{A}}_l)$ and $nnz(\widetilde{\mathbf{B}}_l)$ are $\mathcal{O}(u_{\text{avg}}\frac{nnz(\mathbf{A})}{m})$ and $\mathcal{O}(v_{\text{avg}}\frac{nnz(\mathbf{B})}{n})$. 
Let ${p_l:=\frac{nnz(\widetilde{\mathbf{A}}_l)}{{rs}/{m}}}$ and ${q_l:=\frac{nnz(\widetilde{\mathbf{B}}_l)}{{rt}/{n}}}$. 
Then, it follows that $nnz(\widetilde{\mathbf{C}}_l)$ is ${\mathcal{O}((1-(1-p_l q_l)^r)\frac{st}{mn})} = {\mathcal{O}(p_l q_l \frac{rst}{mn})}={\mathcal{O}(w_{\text{avg}}\frac{nnz(\mathbf{A})nnz(\mathbf{B})}{rmn})}$ for each ${l\in [N]}$. 
Similarly, for each ${l\in [N+1:N+R]}$, $nnz(\widetilde{\mathbf{C}}_l)$ is $\mathcal{O}(w^{*}_{\text{avg}}\frac{nnz(\mathbf{A})nnz(\mathbf{B})}{rmn})$. 
Combining these results, it follows that
$nnz(\mathbf{y}^{(a,b)})$ is $\mathcal{O}((w_{\text{avg}}(N-S)+w^{*}_{\text{avg}}R)\frac{nnz(\mathbf{A})nnz(\mathbf{B})}{rst})$ for each ${a\in [\frac{s}{m}]}$ and each ${b\in [\frac{t}{n}]}$.
Thus, the computational complexity of multiplying the matrix $\mathbf{G}^{+}$ by the vector $\mathbf{y}^{(a,b)}$ is $\mathcal{O}(K(w_{\text{avg}}(N-S)+w^{*}_{\text{avg}}R)\frac{nnz(\mathbf{A})nnz(\mathbf{B})}{rst})$. 
Since such a multiplication must be performed for all ${a\in [\frac{s}{m}]}$ and all ${b\in [\frac{t}{n}]}$, the overall decoding complexity is given by 
\begin{dmath*}
\mathcal{O}\left(K^2(N-S+R)+K(w_{\text{avg}}(K-R)+w^{*}_{\text{avg}}R)\frac{nnz(\mathbf{A})nnz(\mathbf{B})}{rmn}\right).
\end{dmath*}

\subsection{Numerical stability}\label{subsec:NS} 
We measure the numerical stability of an SRKRP code by its average relative error defined as $\eta_{\text{avg}} := \mathbb{E}[\|\mathbf{C}-\widehat{\mathbf{C}}\|_2/ \|\mathbf{C}\|_2]$, where the expectation is taken over a pre-specified distribution of the input matrices $\mathbf{A}$ and  $\mathbf{B}$ and over the distribution of the coding vectors $\mathbf{p}_1,\dots,\mathbf{p}_{N},\mathbf{p}_{N+1},\dots,\mathbf{p}_{N+R}$ and $\mathbf{q}_1,\dots,\mathbf{q}_{N},\mathbf{q}_{N+1},\dots,\mathbf{q}_{N+R}$, for an arbitrary configuration of stragglers. 
While a theoretical analysis of the average relative error of neither SRKRP codes nor the original RKRP codes is currently available, our simulation results---presented in Section~\ref{sec:SIM}---show that even when the coding vectors $\mathbf{p}_1,\dots,\mathbf{p}_N$ and $\mathbf{q}_1,\dots,\mathbf{q}_N$ are very sparse, 
the average relative error of SRKRP codes with only a few dense coding vectors $\mathbf{p}_{N+1},\dots,\mathbf{p}_{N+R}$ and $\mathbf{q}_{N+1},\dots,\mathbf{q}_{N+R}$ is comparable to the average relative error of the original RKRP codes.

\section{Simulation Results}\label{sec:SIM}
In this section, we present the results of our simulations for the failure probability and the numerical stability of SRKRP codes. 
Unless stated otherwise, 
for each set of parameters being considered, we have performed Monte-Carlo simulations until 100 failures were observed (i.e., 100 rank-deficient matrices $\mathbf{G}$ were generated). 

\begin{figure}[t!]\hspace*{-0.25cm}
    \centering
    \includegraphics[scale=0.55]{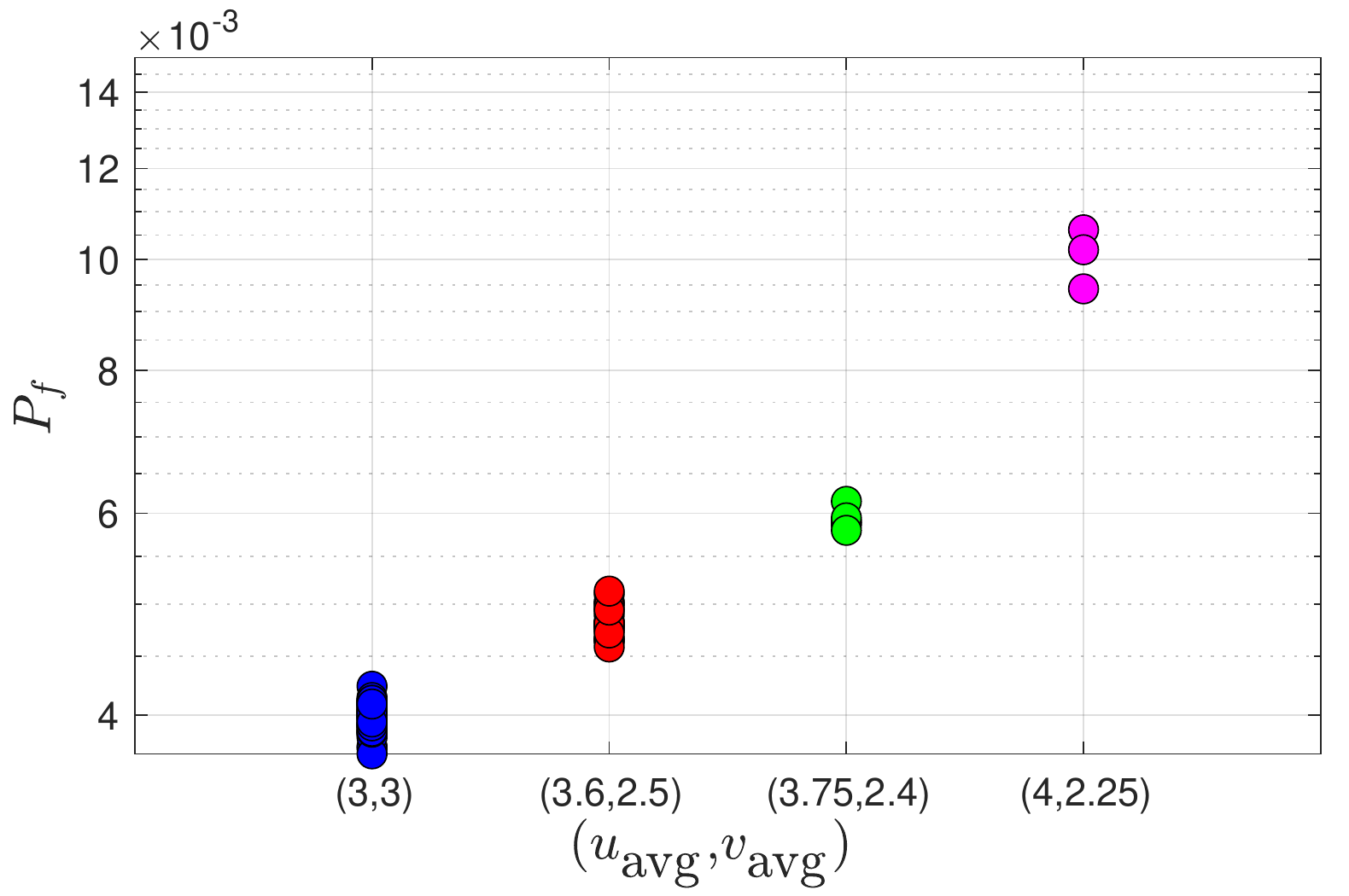}\vspace{-0.165cm}
    \caption{Failure probability ($P_f$) for different weight distribution pairs $(\mathrm{U}(x),\mathrm{V}(x))$ yielding the same overall average weight $w_{\text{avg}}=u_{\text{avg}}v_{\text{avg}}$.}
    \label{fig:plot1}
\end{figure}

Fig.~\ref{fig:plot1} presents the failure probability of SRKRP codes with parameters $m=n=8$ ($K=64$) and ${N-S = K}$, for different pairs of weight distributions \[{\mathrm{U}(x) = \alpha_1 x^2+\beta_1 x^3 + (1-\alpha_1-\beta_1)x^4}\] \[{\mathrm{V}(x) = \alpha_2 x^2+\beta_2 x^3 + (1-\alpha_2-\beta_2)x^4},\] 
for all ${\alpha_i,\beta_i\in \{0,0.05,0.1,\dots,1\}}$ such that $w_{\text{avg}} = u_{\text{avg}}v_{\text{avg}} = (4-2\alpha_1-\beta_1)(4-2\alpha_2-\beta_2)=9$. 
For fixed $(u_{\text{avg}},v_{\text{avg}})$, each point corresponds to the failure probability for a different pair $(\mathrm{U}(x),\mathrm{V}(x))$ with the average weights $u_{\text{avg}}$ and $v_{\text{avg}}$, respectively. 
As can be seen, the failure probability is (almost) the same for different pairs $(\mathrm{U}(x),\mathrm{V}(x))$ with the same $(u_{\text{avg}},v_{\text{avg}})$. 
In addition, it can be seen that the minimum failure probability corresponds to those distribution pairs with $u_{\text{avg}} = v_{\text{avg}}=3$ ($=\sqrt{w_{\text{avg}}}$).

\begin{figure}[t!]\hspace*{-0.35cm}
    \centering
    \includegraphics[scale=0.55]{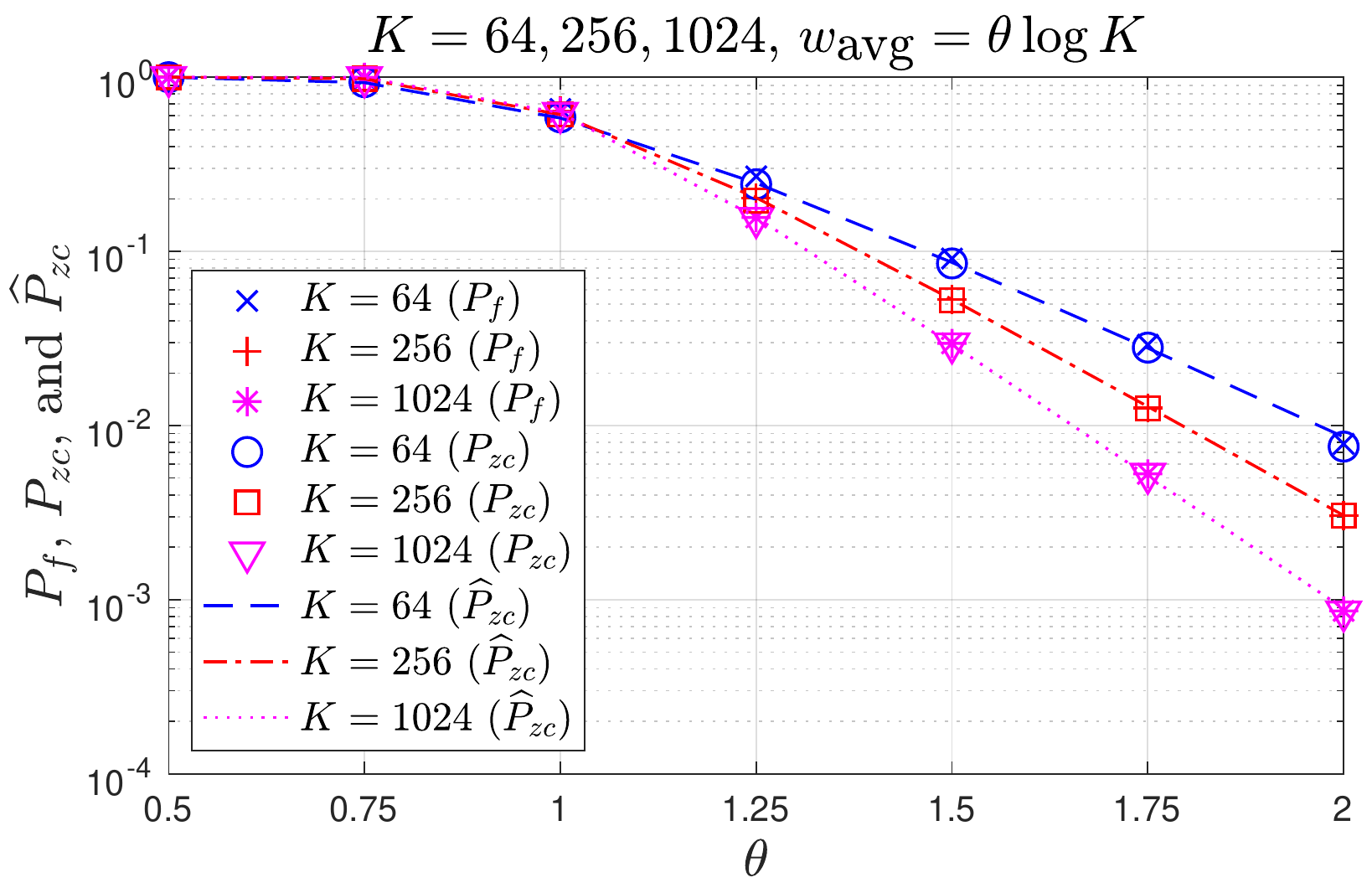}\vspace{-0.125cm}
    \caption{Failure probability ($P_f$), probability of an all-zero column ($P_{zc}$) and its approximation  ($\widehat{P}_{zc}$), for different values of $K$ and $w_{\text{avg}}$.}
    \label{fig:plot23}
\end{figure}  

Fig.~\ref{fig:plot23} depicts the failure probability of SRKRP codes, 
the probability of existence of an all-zero column in the generator matrix of such codes, and 
the probability of existence of an all-zero column in a random binary matrix of the same size and the same sparsity as the generator matrix, 
for parameters ${m=n=8,16,32}$ ($K=64,256,1024$) and ${N-S=K}$, and 
weight distributions $\mathrm{U}(x)$ and $\mathrm{V}(x)$ equal to $ \myLambda(x)$ (as defined in Section~\ref{sec:PA}) with $w_{\text{avg}} = \theta \log K$ for ${\theta\in \{0.5,0.75,1,\dots,2\}}$. 
As can be seen, for fixed $K$, as $\theta$ increases, 
the failure probability decreases, and 
for $\theta>1$, the decay is (almost) exponential in $\theta$. 
In addition, for larger $K$, the failure probability decays faster as $\theta$ increases. 
It can also be seen that the failure probability is very close to (i) the probability of existence of an all-zero column in the generator matrix, and (ii) the probability of existence of an all-zero column in a random binary matrix with the same size and the same sparsity as the generator matrix. 



\begin{figure}[t!]\hspace*{-0.35cm}
    \centering
    \includegraphics[scale=0.55]{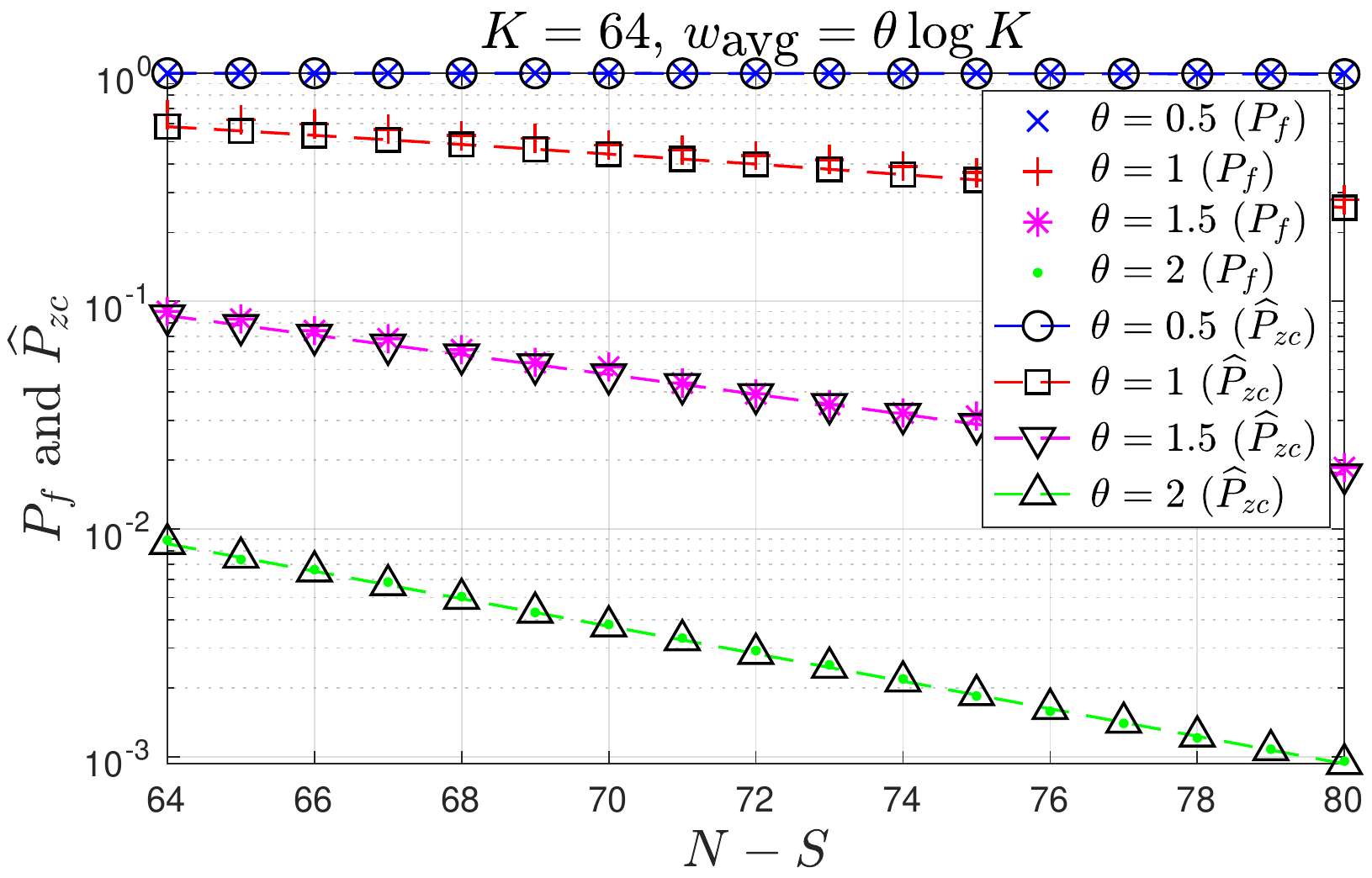}\vspace{-0.05cm}
    \caption{Failure probability ($P_f$) and approximate probability of an all-zero column ($\widehat{P}_{zc}$) for different values of $N-S$ and $w_{\text{avg}}$.}
    \label{fig:plot4}
\end{figure}

Fig.~\ref{fig:plot4} depicts the failure probability and its approximation for parameters $m=n=8$ ($K=64$) and $N-S \in \{K,K+1,\dots,K+16\}$, and weight distribution $\myLambda(x)$ with $w_{\text{avg}} = \theta\log K$ for ${\theta\in \{0.5,1,1.5,2\}}$. 
As can be seen, the failure probability and its approximation are close to each other, not only for $N-S=K$, but also for $N-S>K$.
In addition, one can see that for fixed $K$, the larger is $w_{\text{avg}}$, the more accurate is the approximation. 

\begin{figure}[t!]\hspace*{-0.35cm}
    \centering
    \includegraphics[scale=0.55]{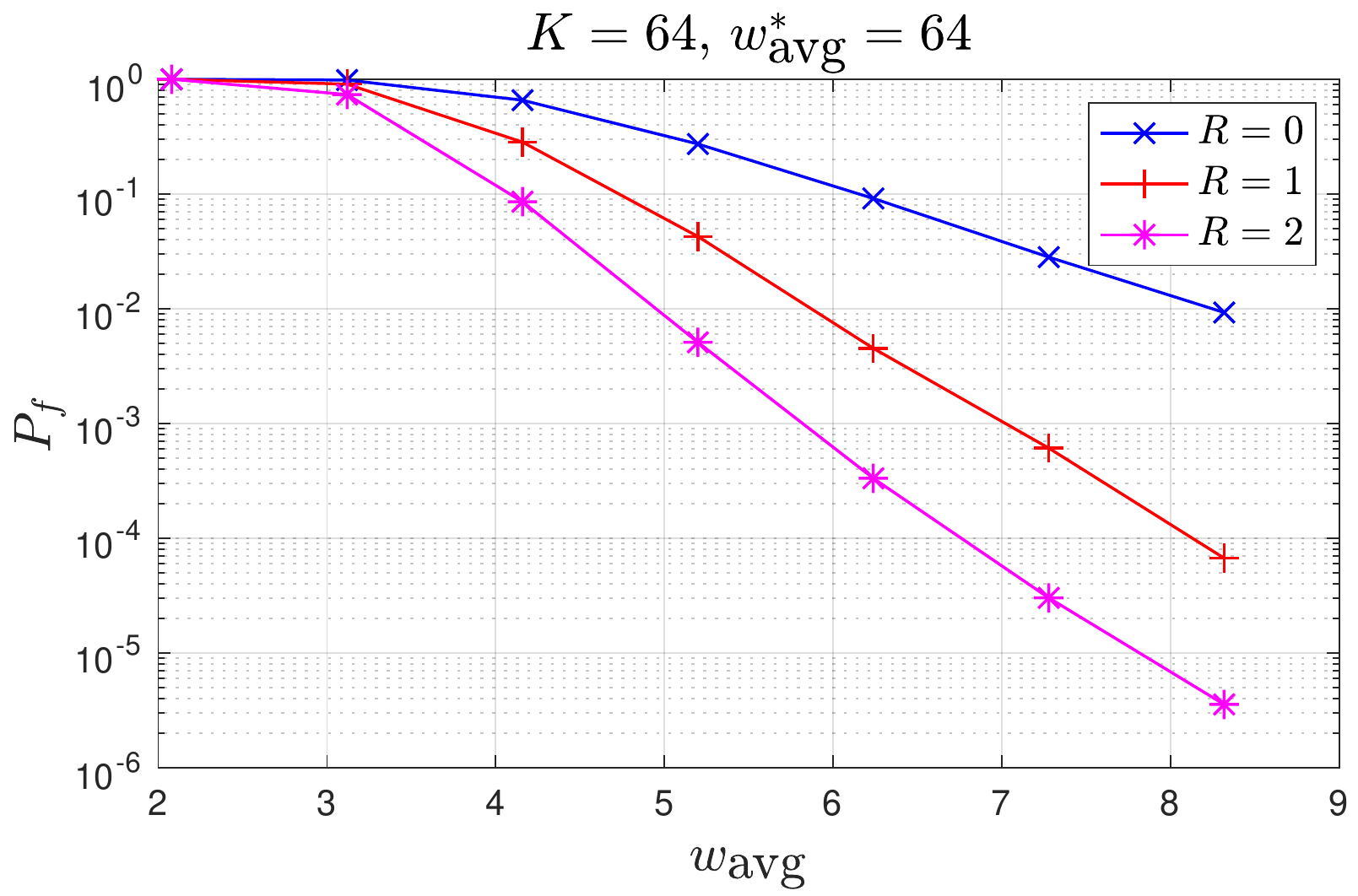}\vspace{-0.125cm}
    \caption{Failure probability ($P_f$) for different values of $R$ and $w_{\text{avg}}$.}
    \label{fig:plot5}
\end{figure}

Fig.~\ref{fig:plot5} depicts the failure probability of SRKRP codes
with ${R=0,1,2}$ dense extra computations performed by the master node (i.e., ${w^{*}_{\text{avg}}=mn}$), 
for parameters ${m=n=8}$ (${K=64}$) and ${N-S=K}$, and 
weight distribution $\myLambda(x)$ with ${w_{\text{avg}} = \theta \log K}$ for ${\theta\in \{0.5,0.75,1,\dots,2\}}$.
As can be seen, the failure probability decays exponentially with $w_{\text{avg}}$, and the decay rate increases linearly with $R$.

\begin{figure}[t!]\hspace*{-0.35cm}
    \centering
    \includegraphics[scale=0.55]{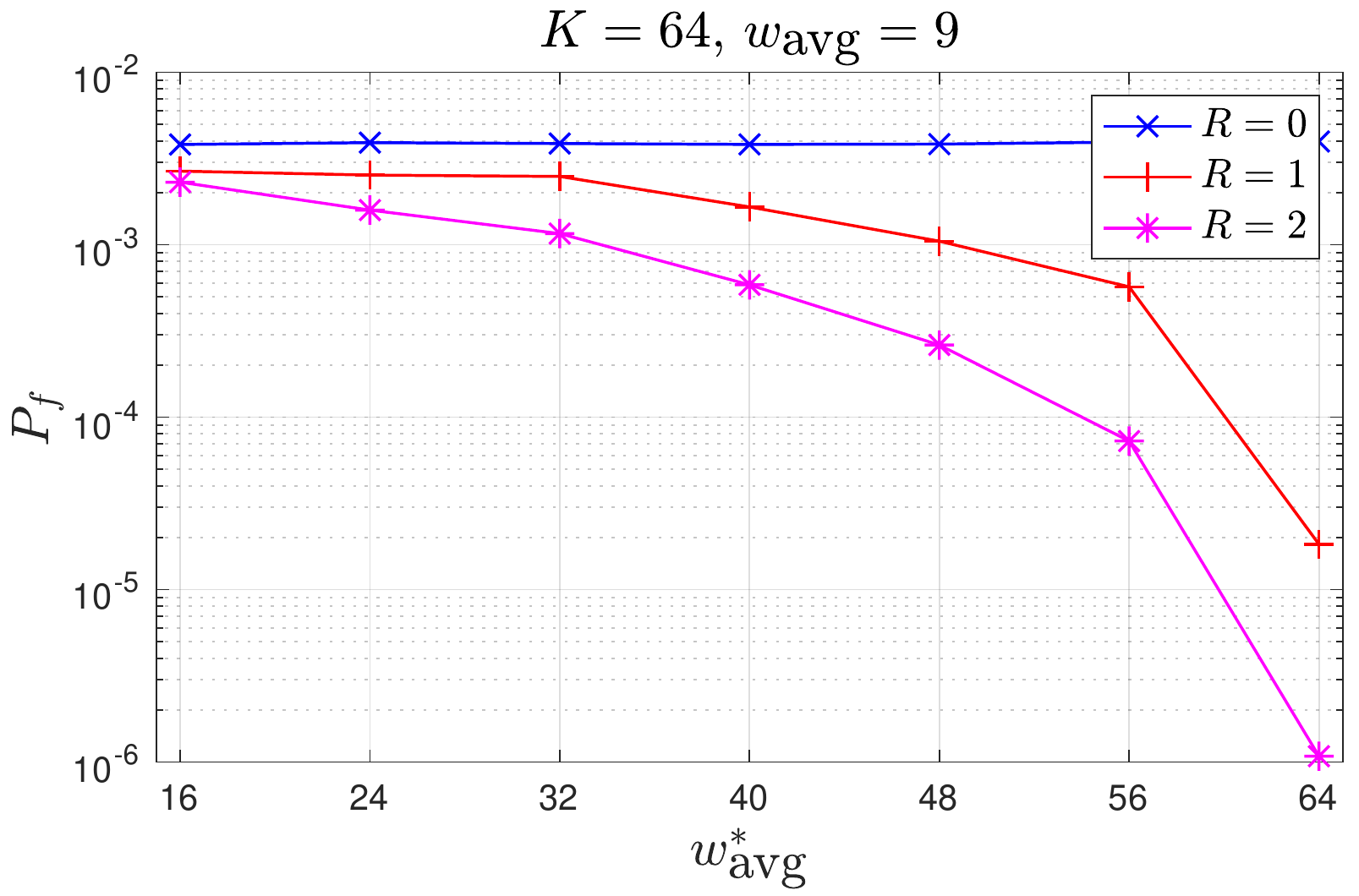}\vspace{-0.125cm}
    \caption{Failure probability ($P_f$) for different values of $R$ and $w^{*}_{\text{avg}}$.}
    \label{fig:plot6}
\end{figure}

For parameters ${m=n=8}$ (${K=64}$) and ${N-S=K}$, and 
weight distribution $\myLambda(x)$ with ${w_{\text{avg}} = 9}$, Fig.~\ref{fig:plot6} depicts the failure probability of SRKRP codes
with ${R=0,1,2}$ extra computations with distributions $\mathrm{U}^{*}(x)=\mathrm{V}^{*}(x)$ equal to $\myLambda^{*}(x)$ (defined similarly as $\myLambda(x)$) for different ${w^{*}_{\text{avg}}\in \{16,24,32,\dots,64\}}$.  
As can be seen, for fixed $R$, the failure probability decreases as $w^{*}_{\text{avg}}$ increases, and for larger $R$, the failure probability decreases faster as $w^{*}_{\text{avg}}$ increases. 

\begin{figure}[t!]\hspace*{-0.35cm}
    \centering
    \includegraphics[scale=0.55]{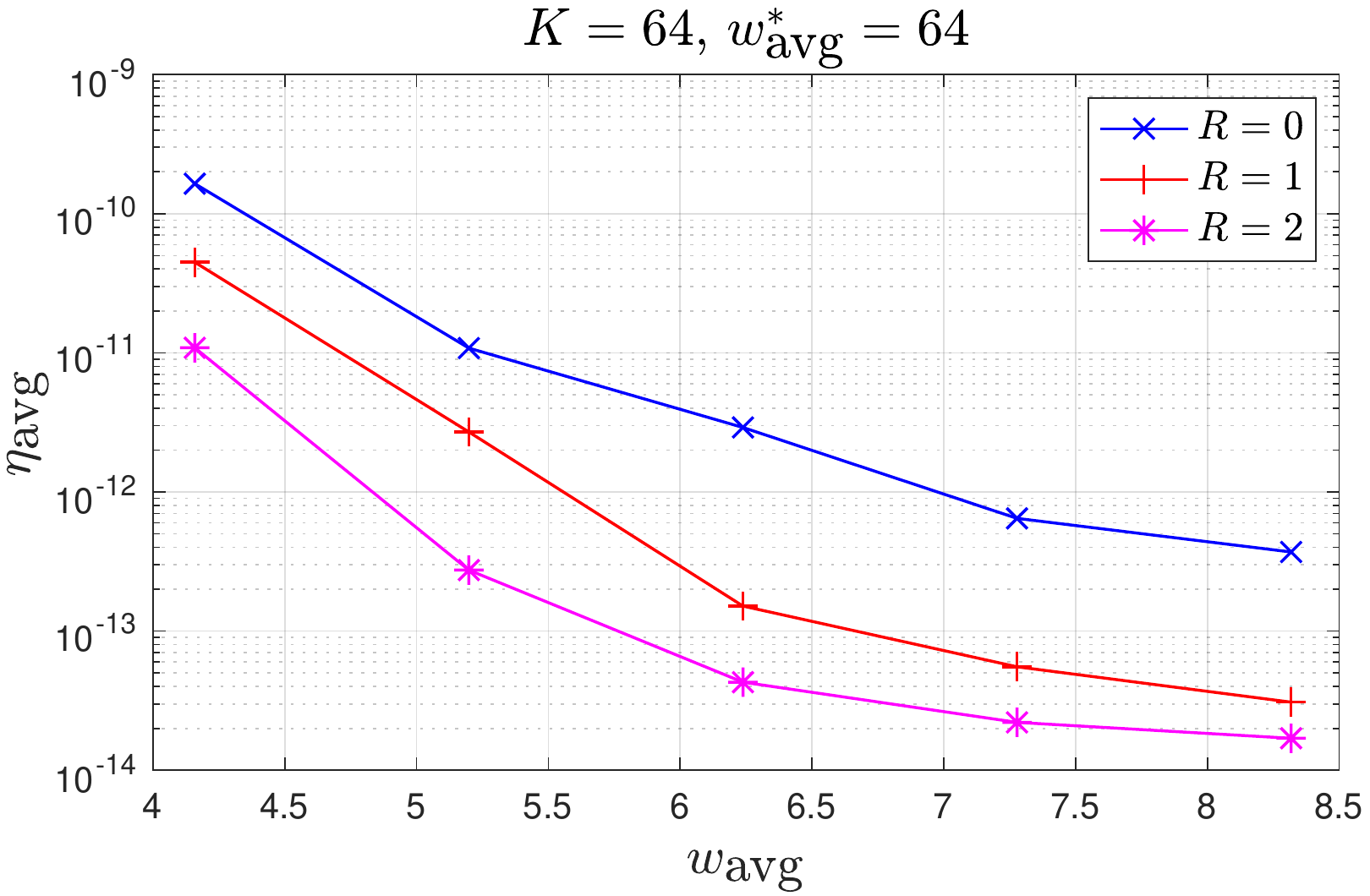}
    \caption{Average relative error ($\eta_{\text{avg}}$) for different values of $R$ and $w_{\text{avg}}$.}
    \label{fig:plot7}
\end{figure}

Fig.~\ref{fig:plot7} depicts the average relative error of SRKRP codes
with ${R=0,1,2}$ dense extra computations, 
for parameters ${m=n=8}$ (${K=64}$) and ${N-S=K}$, and 
weight distribution $\myLambda(x)$ with ${w_{\text{avg}} = \theta \log K}$ for ${\theta\in \{1,1.25,\dots,2\}}$.
For each set of parameters being considered, we have performed $10^6$ Monte-Carlo simulations; and 
for each simulation, we have chosen the entries of the input matrices $\mathbf{A}\in \mathbb{R}^{64\times 64}$ and $\mathbf{B}\in \mathbb{R}^{64\times 64}$ to be realizations of i.i.d.~standard normal random variables. 
As can be seen in Fig.~\ref{fig:plot7}, for fixed $R$, the average relative error decreases as $w_{\text{avg}}$ increases, and for fixed $w_{\text{avg}}$, the average relative error decreases as $R$ increases. 
It should be noted that for an SRKRP code with $R=1$ and $w_{\text{avg}}>7$ (or $R=2$ and $w_{\text{avg}}>6$), the average relative error is between $10^{-13}$ and $10^{-14}$, which is almost the same as the average relative error of a fully dense RKRP code with the same parameters $m,n,K,N-S$ (cf.~\cite{SHN2019}).

\bibliographystyle{IEEEtran}
\bibliography{NewRefs}

\end{document}